\documentclass[twocolumn,aps]{revtex4}
\input epsf.tex
\def\DESepsf(#1 width #2){\epsfxsize=#2 \epsfbox{#1}}
\begin{document}
\title{Large neutrino mixing angles for 
type-I see-saw mechanism in SO(10) GUT  }
\author{Bipin R. Desai}
\affiliation{Physics Department, University of California,
Riverside, CA 92521, USA}
\author{G. Rajasekaran}
\affiliation{Institute of Mathematical Sciences, Chennai 600 113, India}
\author{Utpal Sarkar}
\affiliation{ Physical Research Laboratory, 
Ahmedabad 380 009, India}

\begin{abstract}

We consider the neutrino mixing angles in an SO(10) GUT with
the usual Higgs structure in which neutrino masses are
explained by the type-I see-saw mechanism. The
Dirac-neutrino Yukawa matrix then has a structure 
similar to that of the $u$-quark. We determine the light
neutrino mass matrix through type-I see-saw mechanism using
the experimentally consistent $u$-quark Yukawa 
matrix. We find that large neutrino mixing-sngles 
emerge naturally in this model. 

\end{abstract}
\maketitle

The most natural extensions of the standard model are the
grand unified theories, in which the strong, the weak and
the electromagnetic gauge coupling constants are unified. 
Quarks and leptons also becomes part of one single 
representation of the grand unified group and hence the
quark and lepton mass matrices become related. While this
makes the model predictive, the largeness of the neutrino
mixing angles becomes difficult to explain keeping
the quark mixing angles small. 

We consider here an SO(10) GUT with standard Higgs
structure \cite{min3} and left-right symmetric extension 
\cite{lr} of the standard
electroweak interaction up to an intermediate symmetry 
breaking scale. The quark and lepton Dirac masses come
from the vacuum expectation value ($vev$) of a bi-doublet
Higgs, while the neutrino masses come from the $vev$ of
triplet Higgs scalars. The left-handed neutrino mass
now comes from two sources. There is a {\it direct} 
contribution to the left-handed neutrino mass coming from 
the $vev$ of the left-handed Higgs triplet scalar \cite{typeII}.
This is also
called a type-II see-saw mechanism, since the $vev$ of the
left-handed triplet Higgs is see-saw suppressed by the lepton
number violating scale. There is also the 
usual {\it see-saw} contribution to the left-handed neutrino 
Majorana mass, which is suppressed by the right-handed
neutrino mass (this is also called the type-I see-saw
mechanism) \cite{seesaw}. 

In SO(10) GUTs the Dirac masses of the quarks and leptons
come from the same Yukawa coupling. As a result one expects
that the mixing angles for both the quark and lepton
sectors are similar. Thus, a small quark mixing angle makes
it difficult to accommodate a large neutrino mixing angle
in the case of see-saw neutrino mass. However, in case of 
type-II see-saw neutrino mass it is possible to make the neutrino
mixing angles large in a supersymmetric SO(10) GUT
when the relation $m_b = m_\tau$ is assumed
\cite{min2}. 

In this article we point out that 
in the case of type-I see-saw neutrino mass it is possible
to make the neutrino mixing angle large still keeping the
quark mixing angles small, while at the same time predict
$m_b = m_\tau$.

We shall consider 
an effective scenario, which may emerge from a 
supersymmetric SO(10) GUT. The fermions belong to the
fundamental representation {\bf 16}-plet of Higgs,
which transforms under the Pati-Salam subgroup 
($G_{422} \equiv 
SU(4)_c \times SU(2)_L \times SU(2)_R$) as,
$$ 
\psi_{iL} \equiv {\bf 16 = (4,2,1) + (\bar 4, 1, 2)} .
$$
$i = 1,2,3$ is the generation index. The right-handed fermions 
($\psi_{iR}$) then belong to the conjugate representation, 
$$\psi_{iR} \equiv {\bf \bar{16} = (\bar 4,2,1) + ( 4,1,2)} .$$ 
The generators of the left-right symemtric group 
$G_{3221} \equiv SU(3)_c \times SU(2)_L \times SU(2)_R 
\times U(1)_{B-L}$ are related to the electric charge by
$$ Q = T_{3L} + T_{3R} + {(B-L) \over 2} = T_{3L} + {Y \over 2} .
$$ 
The quarks and leptons belong to the representations
\begin{eqnarray}
(4,2,1) &=& \left\{ \begin{array}{l} 
q_L = \pmatrix{ u&d}_L \equiv (3,2,1,1/3) \cr
\ell_L = \pmatrix{ \nu & e }_L \equiv (1,2,1,-1) \end{array}
\right . \nonumber \\ && \nonumber \\
(\bar 4,1,2) &=& \left\{ \begin{array}{l} 
{q_R}^c = {q^c}_L = \pmatrix{d^c & u^c}_L \equiv (\bar 3,1,2,-1/3) \cr
{\ell_R}^c = {\ell^c}_L = 
\pmatrix{e^c & \nu^c }_L \equiv (1,1,2,1) \end{array} \right . \nonumber \\
\end{eqnarray}
where (x,y,z) and (x,y,z,w) denote the transformation property
under $G_{422}$ and $G_{3221}$ respectively.

The Higgs scalars which are responsible for the electroweak
symmetry breaking and the fermion masses belong to the
representations {\bf 10} (H) and {\bf 126} ($\Delta$).
The $vev$s of the right-handed triplets $\Delta_R \equiv (1,1,3,-2)$
give Majorana masses to the right-handed neutrinos and the
$vev$s of the left-handed triplets $\Delta_L \equiv (1,3,1,-2)$
give tiny Majorana masses to the left-handed neutrinos directly.
The $vev$s of bi-doublet $H$ give Dirac masses to all fermions.
The effective Yukawa couplings are 
\begin{eqnarray}
{\cal L}_Y &=& Y_u \overline{q_L} u_R H + Y_d \overline{q_L}
d_R H^\dagger + Y_n \overline{\ell_L} \nu_R H \nonumber \\ &+&
Y_e \overline{\ell_L} e_R H^\dagger +
f_L \nu_L \nu_L \Delta_L + f_R \nu_R \nu_R \Delta_R.
\end{eqnarray}
The left-handed light neutrino mass matrix is then given by
\[
M_\nu = M_L + M_D^T M_R^{-1} M_D =  M_L + \kappa
\]
where the first term is the direct neutrino mass coming from
the $vev$s of the neutral component of the triplet Higgs scalar
$\Delta_L$. Since the $vev$ of
the scalar $\Delta_L$ acquires a see-saw value of the order of
$<H>^2/<\Delta_R>$, this is
also called the type-II see-saw mechanism. The
second term is the usual or the type-I see-saw neutrino mass.
We adopt the notation $M_D = f <H>$,
$M_L = f_L <\Delta_L>$ and $M_R = f_R <\Delta_R>$. We shall
study the scenario with $M_L < \kappa$, {\it i.e.}, the usual
type-I see-saw neutrino mass dominates over the type-II
see-saw neutrino mass. 

It is wellknown that since the Higgs scalar (H) is in a 
10-plet of SO(10) with fermions in the 16-plet, the structure
of the Dirac-neutrino mass matrix will be the same as that of 
the $u$-quark \cite{min2,14}. 
The structure of the $u$-quark mass matrix at low
energies has been studied extensively \cite{15} and the 
following is obtained for the corresponding Yukawa matrix,
$Y_u$,
\begin{equation}
Y_u \approx \pmatrix{ \lambda^7 &  \lambda^6 &  \lambda^6 \cr
\lambda^6 &  \lambda^4 &  \lambda^4 \cr
\lambda^6 &  \lambda^4 &  1 } m_t
\end{equation}
where $m_t$ is the top-quark mass. The above matrix is usually
written in terms of a parameter $\epsilon$ ($\approx 0.05$), 
but we have expressed it in terms of the more familiar,
Cabbibo parameter $\lambda$ ($\approx .22$), utilizing the 
simple, numerical relation $\epsilon \approx \lambda^2$. A
texture zero is also usually inserted for the 11-element of
$Y_u$, however, we have put a non-zero but very small value
for this matrix element so that the ratio of the diagonal
terms manifestly reproduce the ratio of the up-quark masses.
Furthermore it is known that, because of very small mixing
angles in the quark-system and very strong mass hierarchy,
the scale evolution of $Y_u$ will be negligible \cite{16},
so we will assume the above form to be valid at the GUT scale
as well. From the SO(10) properties mentioned earlier we will then
have, for the Dirac-neutrino,
\begin{equation}
M_D \approx \pmatrix{ \lambda^7 &  \lambda^6 &  \lambda^6 \cr
\lambda^6 &  \lambda^4 &  \lambda^4 \cr
\lambda^6 &  \lambda^4 &  1 } m_3
\end{equation}
where $m_3$ is the value of the 33-component.

For simplicity, we will take the right-handed Majorana
neutrinos mass matrix to be diagonal,
\begin{equation}
M_R = \pmatrix{ M_{11} & 0 & 0 \cr 0 & M_{22} & 0 \cr
0 & 0 & M_{33}}.
\end{equation}
If in addition we assume that this matrix is hierarchical
so that $M_{11} \ll M_{22} \ll M_{33}$ then we can, to a
good approximation, neglect the contribution of $M_{22}$
and $M_{33}$ in the type-I see-saw relation to obtain the
light neutrino mass matrix as
\begin{equation}
\kappa = M_{\nu D}^T M_R^{-1} M_{\nu D} = 
\left( { m_3^2 \over M_{11}} \right) ~ \lambda^{12} ~
\pmatrix{
\lambda^2 & \lambda & \lambda \cr 
\lambda & 1 & 1 \cr \lambda & 1 & 1 } . 
\end{equation}

This matrix shows a maximal angle in the atmospheric neutrino
sector. If we allow for a small correction of order $\lambda$
in the 22-element of $\kappa$ which could be included, for
example, by retaining the matrix elements of $M_R$ in addition to 
just the 11-element, we would obtain
\begin{equation}
\kappa = \left( { m_3^2 \over M_{11}} \right) ~ \lambda^{12} ~
\pmatrix{
\lambda^2 & \lambda & \lambda \cr 
\lambda & 1+\lambda & 1 \cr \lambda & 1 & 1 } . 
\end{equation}
This matrix will reproduce not only the near maximal atmospheric angle
but also the large solar and a small angle in the 13-sector \cite{14,17}.

As for the neutrino masses, the eigenvalues of the above mass
matrix are $(0,\lambda, 2)$, apart from a common multiplicative 
constant, which leads to normal hierarchy, $m_1 \ll m_2 \ll m_3$,
and gives
$$ {\Delta m_{12}^2 \over \Delta_{23}^2} \approx {\lambda^2 \over 4}
= 10^{-2} $$ which is consistent with the data. 
Furthermore, because of the absence of quasi-degeneracy in the masses,
large renormalization group effects in the neutrino sector are not
expected so that the large mixing angles at the GUT scale indicated
by the mass matrix in (7) will stay large at low energies \cite{1a,11}.

We further note that because $H$ in our formalism is a 10-plet,
it predicts $m_b = m_\tau$, which is consistent with experiments
\cite{min2}. The remaining matrix elements in $Y_e$ and $Y_d$ are 
quite small ($\leq \lambda^2$) compared to these masses and any 
discrepancy between them can be explained in terms of a small
contribution of a 126-plet $\Delta$ and by introducing appropriate 
texture zeroes in $Y_e$ and $Y_d$ \cite{11,12}. 

We thus conclude that large neutrino mixing angles are feasible, and
are compatible with small quark mixing angles, in the type-I seesaw
mechanism in a certain class of SO(10) GUT model. We demonstrated
this with a specific realistic example. 

{\sl Acknowledgement}
GR acknowledges the support of the DAE-BRNS Senior Scientist Scheme and
the hospitality of the Physics Department,University of California,
Riverside. BRD acknowledges
the support in part by the U.S. Department of Energy under 
Grant No. DE-FG03-94ER40837.


\end{document}